# Lithography-free IR polarization converters via orthogonal in-plane phonons in $\alpha$-MoO$_3$ flakes

Sina Abedini Dereshgi[1], Thomas G. Folland[2,3], Akshay A. Murthy[4,5], Xianglian Song[1,6], Ibrahim Tanriover[1], Vinayak P. Dravid[4,5,7], Joshua D. Caldwell[2] and Koray Aydin[1,5,*]

[1] Department of Electrical and Computer Engineering, Northwestern University, Evanston, Illinois 60208, United States

[2] Department of Mechanical Engineering, Vanderbilt University, Nashville, Tennessee 37212, USA

[3] Department of Physics and Astronomy, The University of Iowa, Iowa City, Iowa 52242, USA

[4] Department of Materials Science and Engineering, Northwestern University, Evanston, Illinois 60208, USA

[5] International Institute for Nanotechnology, Northwestern University, Evanston, Illinois 60208, USA

[6] International Collaborative Laboratory of 2D Materials for Optoelectronic Science & Technology of Ministry of Education, Engineering Technology Research Center for 2D material Information Function Devices and Systems of Guangdong Province, College of Optoelectronic Engineering, Shenzhen University, Shenzhen 518060, China

[7] Northwestern University Atomic and Nanoscale Characterization Experimental (NU*ANCE*) Center, Northwestern University, Evanston, Illinois 60208, USA

[*] aydin@northwestern.edu

## Abstract.

Exploiting polaritons in natural vdW materials has been successful in achieving extreme light confinement and low-loss optical devices and enabling simplified device integration. Recently, $\alpha$-MoO$_3$ has been reported as a semiconducting biaxial vdW material capable of sustaining naturally orthogonal in-plane phonon polariton modes in IR. In this study, we investigate the polarization-dependent optical characteristics of cavities formed using $\alpha$-MoO$_3$ to extend the degrees of freedom in the design of IR photonic components exploiting the in-plane anisotropy of this material. Polarization-dependent absorption over 80% in a multilayer Fabry-Perot structure with $\alpha$-MoO$_3$ is reported without the need for nanoscale fabrication on the $\alpha$-MoO$_3$. We observe coupling between the $\alpha$-MoO$_3$ optical phonons and the Fabry-Perot cavity resonances. Using cross-polarized reflectance spectroscopy we show that the strong birefringence results in 15% of the total power converted into the orthogonal polarization with respect to incident wave. These findings can open new avenues in the quest for polarization filters and low-loss, integrated planar IR photonics and in dictating polarization control.

**Introduction.**

Birefringence has revolutionized optical systems and has been tailored to many polarization-sensitive applications such as liquid-crystal displays[1], light modulators[2], spectroscopy systems[3], waveplates[4,5,6], nonlinear optical devices[6-8] and biological tissue imaging[9]. Mid-infrared wavelength range encompasses an atmospheric window between 3 to 5 and 8 to 14 μm that is critical for remote sensing applications that require optical components. However, realizing these components in infrared (IR) remains challenging due to the scarcity of material systems that exhibit strong birefringence in this frequency range. Hyperbolicity[10,11] is an extreme form of birefringence where one or more of the crystal axes have a negative dielectric function, whilst the others are positive. Therefore, not only does hyperbolicity address anisotropy, but also, it reduces the required device dimensions by few orders of magnitude[8,12]. Hyperbolicity is also the backbone of exceptional optical phenomena such as hyperlensing[13-15], enhanced thermal radiation[16,17], canalization[11] and negative refraction[18]. Originally hyperbolicity was primarily demonstrated using artificial materials (*aka* metamaterials) consisting of reflective and transparent domains or dielectrics developed with intricate fabrication methods that rely on periodic sub-wavelength features patterned using laborious lithography techniques[12,19-22]. As a result, the geometric quality of the periodic cell and associated fabrication non-idealities, compromises the photonic figures of merit of the devices[23]. Natural hyperbolicity, on the other hand, has been reported in several materials that offer the advantage of relaxing expensive and time-consuming micro/nanofabrication efforts[12,24]. Thus, natural in-plane hyperbolicity offers new possibilities for birefringent optical components and polarization-dependent photonics.

The emergence of van der Waals (vdW) materials and associated ease of integration with on-chip photonics has redirected interest to developing optical components based on them[25]. These materials can be exfoliated down to monolayer thicknesses and transferred onto arbitrary substrates and incorporated into 2D heterostructures[26]. Naturally hyperbolic vdW materials for IR are often studied in the context of polaritons, *i.e.* hybrid light-matter quasi-particles[27]. Hexagonal boron nitride (hBN) is the most widely investigated of this class, featuring two Reststrahlen (RS) bands where hyperbolic phonon polaritons (PhP) can be supported[28,29]. This results from the fact that within these two spectral bands, hBN exhibits negative values of the real part of the permittivity along either the in-plane or out-of-plane directions, while the orthogonal direction(s) exhibit positive values. In-plane anisotropy can be achieved within hBN via fabrication which provides opportunities for applications in polarization manipulation[30]. However, in-plane hyperbolicity is not naturally supported in hBN due to the in-plane symmetry of its crystal lattice. In the context of hyperbolic plasmon polaritons (PPs), graphene and black phosphorus (BP) are the most widely studied materials. While graphene does not support in-plane anisotropy, simulation efforts have demonstrated that introducing anisotropy through asymmetric patterning achieves in-plane birefringence[31]. On the other hand, BP supports weak in-plane anisotropy and hence birefringence[32,33]. In order to enhance anisotropy, BP can be patterned to support anisotropic plasmons which are shown to modify the polarization of the reflected or transmitted beam at IR frequencies[34]. However, adding a substrate significantly deteriorates the amount of polarization rotation as well as the intensity of the output, especially for the reflected beam[33]. In order to boost the in-plane anisotropy and ameliorate the effect of the substrate, BP was considered in the free-standing form, rendering such devices impractical. Besides, in case of patterned vdW materials, due to the fast scattering times associated with light-electron coupling, (Ohmic) losses are inherently high for PPs compared to PhPs[10]. Recent works demonstrated that in-plane hyperbolicity is naturally present in the highly anisotropic vdW 2D material $\alpha$-MoO$_3$, featuring three distinct RS bands, resulting in a variety of different hyperbolic behaviors in the long-wave IR[36,37]. The low-loss PhPs supported by MoO$_3$ offer a platform to better control polarization in this spectral range. Here, we experimentally show that the naturally in-plane hyperbolicity of $\alpha$-MoO$_3$ supports polarization-dependent resonant absorption (polarization filter) by optical phonons (OPh) and polarization conversion in the mid-infrared when integrated into a FP cavity. Unlike PhPs that require momentum-matching techniques, OPhs require low momenta and can be directly excited with photons. As a result, the necessity of patterning with lithography techniques is obviated.

$\alpha$-MoO$_3$ has garnered much attention as a biaxial vdW material, due to its polar and highly anisotropic optical phonons within the mid-infrared (MIR) that offers extra RS bands compared to hBN[36,37]. In light of this discovery, a few publications have experimentally mapped the dispersion relation of flakes using near-field and far-field optical microscopies via scattering-type scanning near-field optical microscopy (s-SNOM) and Fourier transform infrared (FTIR) spectroscopy, respectively[36,37]. The strong anisotropy in $\alpha$-MoO$_3$, originating from lattice anisotropy, gives rise to three distinct lattice modes along the three orthogonal principle crystal directions leading to the previously reported naturally occurring in-plane hyperbolicity. $\alpha$-MoO$_3$ is an orthorhombic crystal featuring three different symmetries for oxygen atoms that give rise to the biaxial optical properties (Fig. 1a)[36,40]. As is the case for all polar crystals, these RS bands are bracketed by longitudinal (LO) and transverse (TO) optic phonons in each direction. The complex permittivity of $\alpha$-MoO$_3$ is illustrated in Figs. 1b and c, which outline the three successive RS bands in directions identified in Fig. 1a as *x, y* and *z* (see model below). Since $\alpha$-MoO$_3$ can be exfoliated along the

[010] crystal direction, this is dubbed the z direction (optical axis). Following the established convention of crystallographic directions in the literature, the x and y directions represent the [100] and [001] directions, respectively[36]. The strong anisotropy of α-MoO$_3$ could be impactful for a wealth of applications, ranging from IR waveplates, polarization sensors, in-plane imaging and hyperlenses to in-plane chemical sensing[38]. As a semiconductor, it can also enable a number of active nanophotonic concepts, while realizing heterostructures with other 2D vdW materials could pave the way to more versatile novel hybrid polaritons[41].

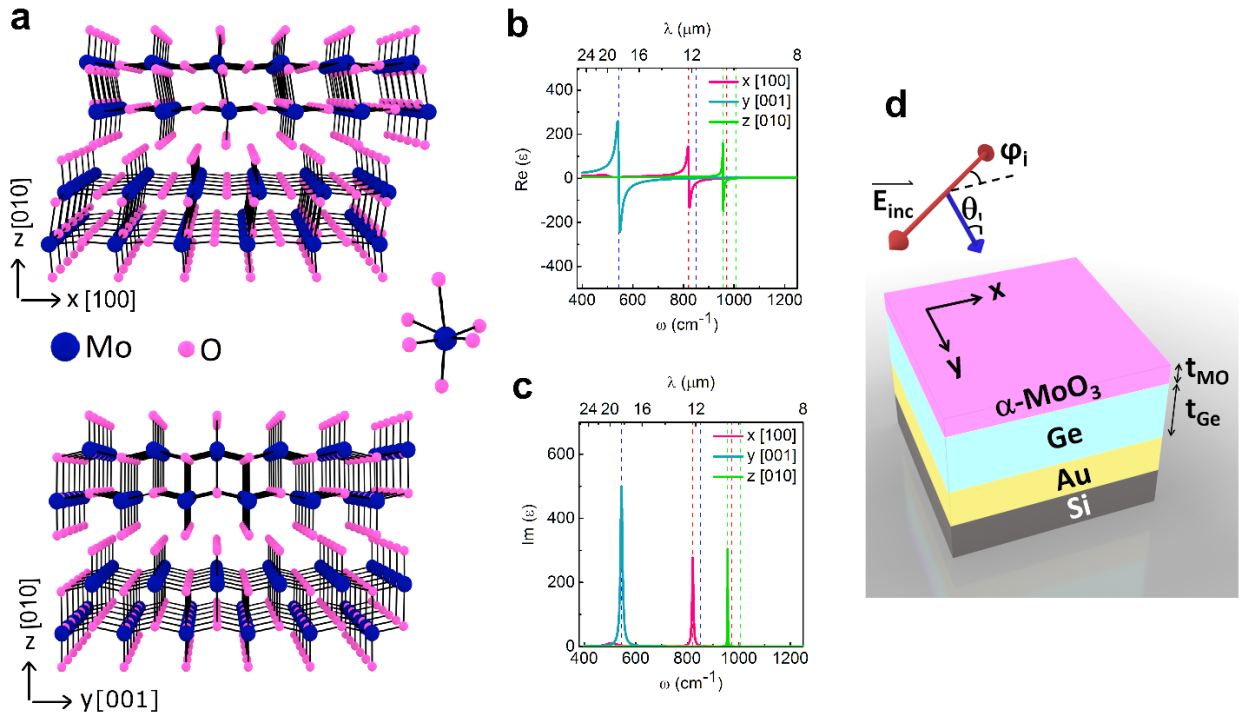

**Figure 1|Optical characteristics of α-MoO$_3$ and the structure under study.** (a) Schematic representation of atomic orientation in the bulk structure of α-MoO$_3$ in xz and yz planes, (b) real and (c) imaginary parts of the dielectric function for α-MoO$_3$. (d) Schematic illustration of the investigated multilayer structure.

In this study, we have carried out experiments that highlight some of the routes that can be investigated in more depth toward realizing large number of degrees of freedom and in-plane nanophotonics. Here, we use the polarization-dependent OPh modes of α-MoO$_3$ to create a Fabry-Perot (FP) perfect absorber geometry, as illustrated in Fig. 1d. As an introduction, first, we model the dielectric function tensor of α-MoO$_3$ to be used for transfer matrix method (TMM) and FDTD simulations[39]. We observe coupling of the FP and OPh modes in this system in simulations and experiments, indicative of strong coupling. The in-plane anisotropy of α-MoO$_3$ has critical implications for planar, on-chip photonics and we demonstrate polarization filters (absorbers) and enhanced polarization conversion resulting from the multiple spectrally overlapping RS bands with simple, proof-of-concept experiments and supporting simulations.

**Results.**

**Optical model of α-MoO$_3$.** Since OPhs dominate the optical characteristics of α-MoO$_3$ in the mid-infrared, the complex permittivity can be described by phenomenological Lorentz function

$$\varepsilon = \sum_{j=x,y,z} \varepsilon_j \cdot \hat{j}, \quad \varepsilon_j = \varepsilon_{\infty,j} \left( \frac{\omega_{LO,j}^2 - \omega^2 - i\omega\Gamma_j}{\omega_{TO,j}^2 - \omega^2 - i\omega\Gamma_j} \right), \tag{1}$$

where $\varepsilon_{\infty,j}$ and $\Gamma_j$ respectively stand for the high frequency permittivity and phonon damping (broadening factor of Lorentz function), while $\omega_{LO,j}$ and $\omega_{TO,j}$ represent LO and TO phonon frequencies that bracket each RS band along the corresponding direction $j$. The values used for these parameters are taken from the recently reported experimental data by Álvarez-Pérez et. al.[39] and the results are provided in Figs. 1b and c. It is evident from this figure that there are three RS bands between 540 and 1010 cm$^{-1}$ (9.9 and 18.5 μm), each of which is designated by vertical color-coded dashed lines in Figs. 1b and c. As a result, there are various hyperbolic regions in this spectral range. For frequencies above $\omega$ = 1007 cm$^{-1}$ (wavelengths below $\lambda$ = 9.93 μm), the real part of permittivity along all directions is positive and unequal. At lower frequencies where 963 cm$^{-1}$ < $\omega$ < 1007 cm$^{-1}$, Re($\varepsilon_z$) < 0 and Re($\varepsilon_x$)Re($\varepsilon_z$) > 0 and Re($\varepsilon_y$) > 0, resulting in type 1 hyperbolicity in $xz$ and $yz$ planes, as shown in prior works[36-39]. At 547 cm$^{-1}$ < $\omega$ < 963 cm$^{-1}$ there are several bands of type 2 hyperbolicities, Re($\varepsilon_z$) > 0 and/or Re($\varepsilon_x$)Re($\varepsilon_z$) < 0, Re($\varepsilon_y$)Re($\varepsilon_z$) < 0. More importantly, when 545 cm$^{-1}$ < $\omega$ < 821 cm$^{-1}$ we observe in-plane hyperbolicity that is essential for manipulating the polarization state of light impinging in the direction of optical axis ($z$).

**Polarization-dependent absorption in α-MoO$_3$.** As a simple geometry to exploit the polarization-dependent properties of $α$-MoO$_3$, we consider a FP structure consisting of an α-MoO$_3$ flake, a Ge cavity, and a gold back plane, illustrated in Fig. 1d. Ge is chosen to form the cavity due to its infrared transparency and lack of significant optical dispersion or loss within the frequency range of interest[41]. The absorption in the structure (equivalent to that of $α$-MoO$_3$, $A_{MO}$) is simply $A = A_{MO}$ = 1 - $R$ where $A$ and $R$ represent absorbance and reflectance respectively. The polarization of the impinging radiation ($\phi_i$) is defined with respect to the $x$-axis throughout this study. The impact of the Ge thicknesses ($t_{Ge}$) on the spectral absorption of the FP structure under *p*-polarized incident light is illustrated in the simulation results presented in Fig. 2.

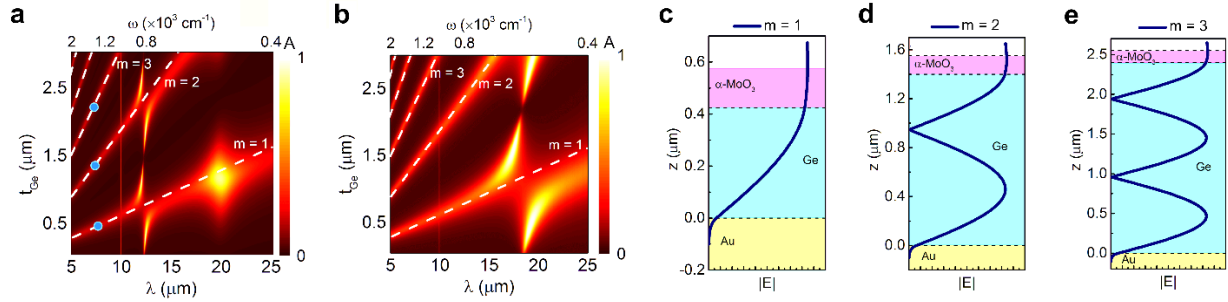

**Figure 2|Dispersion of α-MoO₃ FP cavity with *t*_Ge*.** (a) Simulated dependence of total absorption on wavelength/frequency and Ge thickness ($t_{Ge}$) where α-MoO₃ thickness is kept constant at $t_{MO}$ = 0.15 μm for (a) *x* ($\phi_i$ = 0°) and (b) *y* ($\phi_i$ = 90°) phonons. Simulated electric field magnitude |**E**| in the cross section of the studied multilayer system for $t_{MO}$ = 0.15 μm and (c) $t_{Ge}$ = 0.42 μm (m = 1), (d) $t_{Ge}$ = 1.4 μm (m = 2) and (e) $t_{Ge}$ = 2.4 μm (m = 3). Note that the scales in panels (c) - (e) are not the same due to the $t_{Ge}$ difference. The dashed lines in panels (a) and (b) represent the FP mode orders. The dots on panel (a) mark the mode profile simulations of panels (c) - (e).

In an effort to gain the requisite insight into the spectral dependence of the cavity absorption on the Ge layer thickness, $t_{Ge}$, we simulated the structure using the TMM method (identical FDTD results with Lumerical in Supplementary Note 1). The results for *x*- optical phonons (OPh$_x$) and y-direction optical phonons (OPh$_y$) are respectively provided in Figs. 2a and b. It can be inferred from these two panels that as $t_{Ge}$ increases, the absorption ascribed to FP modes (white dashed lines) redshifts up to a certain thickness, they overlap with OPh$_x$ (OPh$_y$) and an anti-crossing between the OPh and FP modes occurs. There are several higher-order FP modes that are marked by dashed lines that show similar behavior as that of the fundamental FP mode (*m* = 1). In Figs. 2a and b oblique lines with steeper slopes represent higher-order FP cavity modes (more details in Supplementary Note 1). In order to verify the FP modes, electric field magnitude (|**E**|) simulations are provided in Figs. 2c-e. A hypothetical vertical line at $\omega$ = 1300 cm$^{-1}$ ($\lambda$ ≈ 8 μm) in Fig. 2a or b would pass through 3 different modes marked on Fig. 2a with blue dots. These dots represent *m* = 1, 2 and 3 FP modes respectively at three different $t_{Ge}$ values, $t_{Ge}$ = 0.42, 1.4 and 2.4 μm. The simulated |**E**| for the mentioned $t_{Ge}$ values in the cross-section of the multilayer structure is illustrated in Figs. 2c-e that verify the mode number engraved in the number of nodes within Ge layer. The electric field is similar for *xz* and *yz* cross sections. Also evident from Figs. 2a and b are the *x* and *y* phonons respectively at 820 cm$^{-1}$ (12.2 μm) and 550 cm$^{-1}$ (18.2 μm) as well as their forbidden-crossing with the discussed cavity modes. The vertical absorption lines observed in the vicinity of 10 μm ($\omega$ = 1000 cm$^{-1}$) in Figs. 2a and b represent the OPh$_z$ modes, which can be stimulated with off-normal excitation. For the purpose of our simulations this was achieved by setting the incident angle ($\theta$) to 25°. This value of the incident angle is defined by the weighted average incident angle of the Cassegrain objectives used for FTIR measurements and OPh$_z$ resonances are threfore expected to be excited in our experiments. Unlike OPh$_x$ and OPh$_y$, OPh$_z$ modes are orthogonal to the FP modes and thus they can be supported at the same frequency without strong modal interactions (e.g. anti-crossing behavior).

In order to observe the traces of OPh modes and distinguish them from the FP cavity modes in reflection (or absorption) results, two values of $t_{Ge}$, 0.42 and 0.85 μm, are identified from simulation results of Fig. 2a. For $t_{Ge}$ = 0.42 μm, absorption based on OPh modes is expected as opposed to $t_{Ge}$ = 0.85 μm, where coupling of OPh modes to the FP mode $m$ = 1 is anticipated (consider two horizontal line cuts from Fig. 2a at the mentioned $t_{Ge}$ values). At these two values of $t_{Ge}$, we also simulate the spectral characteristics of absorption in $\alpha$-MoO$_3$ as a function of $t_{MO}$ in Figs. 3a and b when $\phi_i$ = 0°. An important take-away from Figs. 3a and b simulations is the increased number of supported OPh modes in $\alpha$-MoO$_3$ for larger $t_{MO}$ values[3]. As $t_{MO}$ increases in Fig. 3a, higher-order modes of OPhs also become accessible. At $\omega$ > 1000 cm$^{-1}$ ($\lambda$ < 10 μm) since $\alpha$-MoO$_3$ is highly dispersive, linear dependence of spectral absorption on $t_{MO}$ is not expected. The oblique lines in this frequency range are FP-like (similar to Fig. 2a and b) which are a result of cavity formed by air-MoO$_3$-Ge layers. As $t_{MO}$ increases, these lines become increasingly non-linear as they approach the RS bands. This is expected since the onset of RS bands is concomitant to substantial changes in the dielectric function, namely the vicinity of $\omega_{LO,z}$, $\omega_{LO,x}$ and $\omega_{LO,y}$. When 800 cm$^{-1}$ < $\omega$ < 1000 cm$^{-1}$ FP modes in $\alpha$-MoO$_3$ cannot be supported in $xz$ cross-section since the real part of the permittivity is negative in this range. Contrarily, for $\omega$ < 800 cm$^{-1}$, FP mode formed by air-MoO$_3$-Ge overlaps with $m$ = 1 OPh$_x$ mode and the hybrid mode redshifts and higher-order OPh$_x$ appears at 820 cm$^{-1}$. The hybrid mode with fundamental OPh$_x$ and higher order OPh$_x$ mode are marked with purple dots in Fig. 3a and their corresponding |**E**| profiles are illustrated in Figs. 3c and d. In these figures, field profiles for the $m$ = 1 and $m$ = 2 when $t_{MO}$ = 0.95 μm and $t_{Ge}$ = 0.42 μm at $\omega$ = 660 cm$^{-1}$ ($\lambda \approx$ 15 μm) and $\omega$ = 805 cm$^{-1}$ ($\lambda \approx$ 10 μm) are presented in Fig. 3c and d, respectively, that support the presented mode discussion. Contrarily, in Fig. 3b where $t_{Ge}$ = 0.85 μm, Ge is thick enough to support FP modes formed by MoO$_3$-Ge-Au layers at $\omega$ < 820 cm$^{-1}$ ($\lambda$ > 12 μm). Consequently, the $m$ = 1 FP mode of Ge also interacts with the prior overlapping modes, $m$ = 1 OPh$_x$ and $m$ = 1 FP of $\alpha$-MoO$_3$. As a result of the three interacting modes (instead of two in the case of Fig. 3a), an additional extinction peak appears near $\omega$ < 900 cm$^{-1}$ which shows opposite trend to that of redshifting hybridized $m = 1$ OPh$_x$ and $m = 1$ FP of α-MoO$_3$. The splitting observed around 18 μm is due to the OPh$_x$ mode near this wavelength. At $xz$ cross-section, since α-MoO$_3$ is highly dispersive and has negative real permittivity values at fequencies between $\omega_{LO,z}$ and $\omega_{TO,x}$, the FP mode of α-MoO$_3$ is not present in this range. Contrarily, at $\omega < \omega_{TO,x}$ for $xz$ cross-section ($\phi_i = 0°$), α-MoO$_3$ has positive real permittivity values and its corresponding FP mode is feasible and it hybridizes with OPh$_x$ modes. For $yz$ cross-section ($\phi_i = 90°$), similar behavior is expected except that the FP mode for α-MoO$_3$ is supported at $\omega < \omega_{TO,y}$ (Supplementary Note 1).

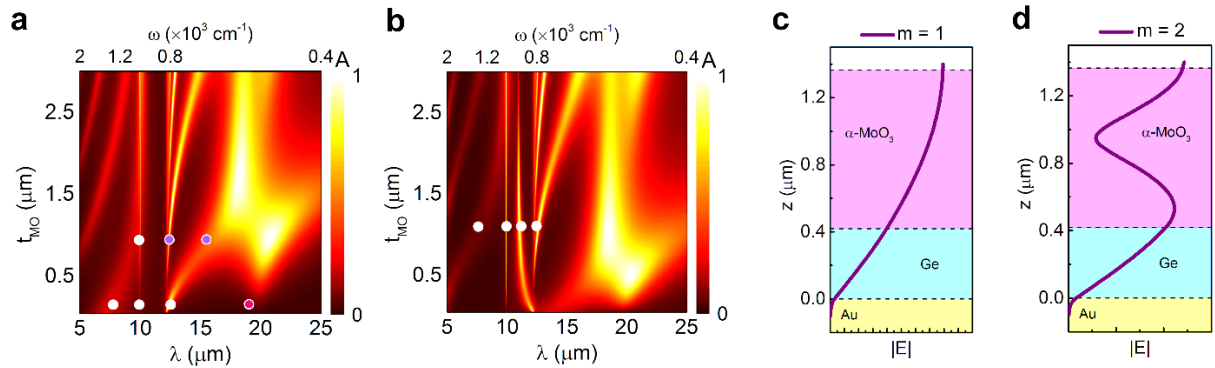

**Figure 3| Dispersion of α-MoO$_3$ FP cavity with $t_{MO}$.** Simulated absorption versus wavelength/frequency and α-MoO$_3$ thickness ($t_{MO}$) in the multilayer structure for constant Ge thickness (c) $t_{Ge}$ = 0.42 μm and (d) $t_{Ge}$ = 0.85 μm where $\phi_i = 0°$. Simulated electric field magnitude |**E**| in the cross section of the studied multilayer system for $t_{MO}$ = 0.15 μm and $t_{Ge}$ = 0.42 μm at (c) $\omega$ = 660 cm$^{-1}$ ($m$ = 1) and (d) $\omega$ = 805 cm$^{-1}$ ($m$ = 2). The dots in panels (a) and (b) mark the peaks in the absorption spectra of the fabricated samples, S1, S2 and S3 and the purple dots also represent the data points for the simulated field profiles in panels (c) and (d). The pink dot is the OPh$_y$ mode which is excited when incident polarization $\phi_i$ = 90° (Supplementary Note 1).

In order to verify our predictions of the coupled FP-OPh modes in α-MoO$_3$ experimentally, three samples like the one schematically illustrated in Fig. 1d were fabricated; sample S1 with $t_{Ge}$ = 0.42 μm and $t_{MO}$ = 0.15 μm, sample S2 with $t_{Ge}$ = 0.42 μm and $t_{MO}$ = 0.95 μm and sample S3 with $t_{Ge}$ = 0.85 μm and $t_{MO}$ = 1.1 μm. The spectral absorption measurements of these samples are presented in Figs. 4a-c, respectively. Details of the fabrication and measurement methods are based on previous reports[28] and are provided in the Methods as well as in Supplementary Note 2 [42,43].

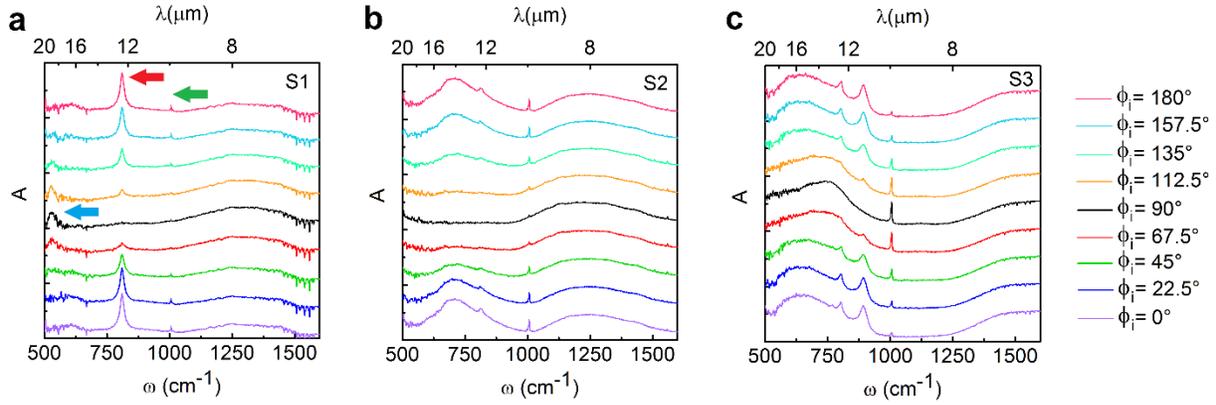

**Figure 4 | Experimental polarization-dependent absorption spectra of α-MoO₃.** Polarization-dependent absorption spectra of fabricated samples with different Ge and α-MoO$_3$ thickness ($t_{Ge}$ and $t_{MO}$, respectively), schematically illustrated in Fig. 1d with thick Au, (a) S1, $t_{Ge}$ = 0.42 μm and $t_{MO}$ = 0.15 μm, (b) S2, $t_{Ge}$ = 0.42 μm and $t_{MO}$ = 0.95 μm, and (c) S3, $t_{Ge}$ = 0.85 μm and $t_{MO}$ = 1.1 μm. Incident polarization ($\phi_i$) is varied (bottom to top) from 0° to 180° in 22.5° revolution and illumination angle $\theta$ =25° from FTIR. The green, red and blue arrows in (a) emphasize respectively the OPh$_z$, OPh$_x$ and OPh$_y$. The peaks observed in each sample are marked on Figs. 3a and b.

Our FTIR spectra show that sample S1 ($t_{Ge}$ = 0.42 μm and $t_{MO}$ = 0.15 μm) supports polarization dependent absorption, as shown in Fig. 4a as a function of the incident polarizations between 0° to 180° in steps of 22.5°. In this figure, the peak at 800 cm$^{-1}$ (12.5 μm) results from absorption from OPh$_x$ in α-MoO$_3$, due to its presence at 0 and 180°. At $\phi_i$ = 90°, another peak in the vicinity of 550 cm$^{-1}$ becomes apparent, corresponding to OPh$_y$ (Fig. 4a highlighted with blue arrow). The absorption peak at 1000 cm$^{-1}$ originates from OPh$_z$ excited as a result of the off-normal incident angle of the Cassegrain objective lens of FTIR. Finally, the broad peak located at a frequency higher than the *z*-LO phonons corresponds to the fundamental FP mode. The absorption tied to OPh$_x$ reaches over 80% for $\phi_i$ = 0° (Supplementary Note 2) and the corresponding full width at half maximum (FWHM) is 26 cm$^{-1}$ (0.4 μm) centered at $\omega$ = 811 cm$^{-1}$. These results are consistent with values observed from the other two samples as shown in Figs. 4b and c. In Fig. 4b (sample S2 with $t_{Ge}$ = 0.42 μm and $t_{MO}$ = 0.95 μm) we observe two peaks in the RS band near 750 cm$^{-1}$; the narrow one at 811 cm$^{-1}$ is the higher order mode of the OPh$_x$ (*m* = 2) and the broader one is the hybridized *m* = 1 OPh$_x$ and *m* = 1 FP of α-MoO$_3$ that redshifts and broadens as $t_{MO}$ increases. These two mode values are marked on Fig. 3a (purple dots) and the corresponding field profiles outlining *m* = 1 and 2 are put forth in Figs. 3c and d. The strong polarization dependence of these peaks confirms that they cannot be attributed to the Ge cavity, which due to its isotropic, cubic crystal structure exhibits no polarization dependence. Finally, the thickness parameters for the sample S3 ($t_{Ge}$ = 0.85 μm and $t_{MO}$ = 1.1 μm) is designed so that the fundamental cavity mode is red-shifted to longer wavelengths (lower frequency), to the vicinity of OPh$_x$ modes. As a result, the $m$ = 1 FP mode overlaps with $m$ = 1 OPh$_x$ and m = 1 α-MoO$_3$ FP modes and strong coupling is evident by the splitting that occurs in the absorption spectra of Fig. 4c. Thus, two distinct peaks at 811 and 720 cm$^{-1}$ are present for the three coupled modes compared to only two for sample S2, which is in line with our predictions of the modal overlap. Finally, the dots on simulation results of Figs. 3a and b mark the observed peaks in experimental results (Figs. 4a-c) and they agree well with simulations. The experimental results demonstrate strong polarization-dependent absorption; Fig. 4a represents 80% absorption and our simulations estimate 100% in the absence of experiment non-idealities. From a more general perspective, Fig. 4c indicates that one can tune absorption to to over 70% (100% in simulations) approximately at 900 and 820 cm$^{-1}$ demonstrating dual-band polarization-dependent perfect absorption action. Due to the flake roughnesses and small flake dimensions compared to the wavelength of test, the experimental results exhibit lower absorption peaks than simulations. The results presented in this section are also evidences of strong birefringence observed by polarization-sensitive absorption and filtering action, particularly at $\omega_{TO,x}$. The polarization filtering contrast, defined as $\Delta R = R(\phi_l = 90) - R(\phi_l = 0)$, is $\Delta R \approx 70\%$ for Sample S1 at 811 cm$^{-1}$. Benefitting modal overlap engineering by tuning the thickness of α-MoO$_3$ and Ge layers, polarization filtering can be realized at several frequency values other than $\omega_{TO,x}$, particularly at 705 and 892 cm$^{-1}$ for samples S2 and S3 respectively. Theoretical values for filtering efficiency are above 90% for all of the absorption peaks.

A few discrepancies are evident in the experimental results compared to FDTD simulations. For all the samples (Figs. 4a-c), OPh$_z$ resonances distinctly reside close to 1000 cm$^{-1}$ and show wide variations in intensity. We attribute this to the roughness of $\alpha$-MoO$_3$ flakes (Supplementary Note 2) as well variations in the high frequency dielectric constant, which will result in some difference in intensity as the polarization state is rotated. The rough surface acts to diffract the radiation, which in turn gives rise to **E** components aligned along the optical axis (z) to excite OPh$_z$. In Fig. 4c, the OPh$_z$ absorption peak at 1006 cm$^{-1}$ (9.94 μm) reaches to 53% and the FWHM and quality factor are 8.5 cm$^{-1}$ (0.084 μm) and 119, respectively. The higher intensity of OPh$_z$ peaks for samples S2 and S3 compared to S1 are due to the thicker $\alpha$-MoO$_3$ flakes in these samples. Another major consequence of roughness is the suppression of OPh$_y$ in samples S2 and S3 in all samples. In order for this mode to be observed, a flake should have negligible roughness in comparison to 18 μm free-space wavelength of the OPh$_y$. The largest flake dimensions are on the order of 30 to 40 μm, which together with the surface roughness of the $\alpha$-MoO$_3$ flakes studied, hinder the observation of clear, high quality $OPh_y$ and lower frequency $OPh_x$ signals (Supplementary Note 2).

**Enhanced birefringence from in-plane RS bands.** A crucial outcome of in-plane anisotropic OPhs is strong intrinsic birefringence. Whilst $\omega_{TO}$ frequencies were tailored to absorbers and polarization filters in the previous section, off-resonance frequencies ($\omega \neq \omega_{TO}$) where absorption drops, are suitable for waveplates and polarization conversion applications. The multiple RS bands of $\alpha$-MoO$_3$ gives rise to unique polarization rotation capabilities for IR photonics. In order to demonstrate its potential, we have carried out a simple, proof-of-concept experiment as illustrated in Fig. 5a, on sample S2. In the FTIR experiment, one linear polarizer is placed on the path of incident beam (polarizer) and one in the path of the reflected beam, before the detector (analyzer). These linear polarization angles are defined with respect to x-axis and are represented with $\phi_i$ and $\phi_a$, respectively, for polarizer and analyzer (Fig. 5a). The experimental reflectance results for $\phi_i$ = 45° and different values of $\phi_a$ is provided in Fig. 5b.

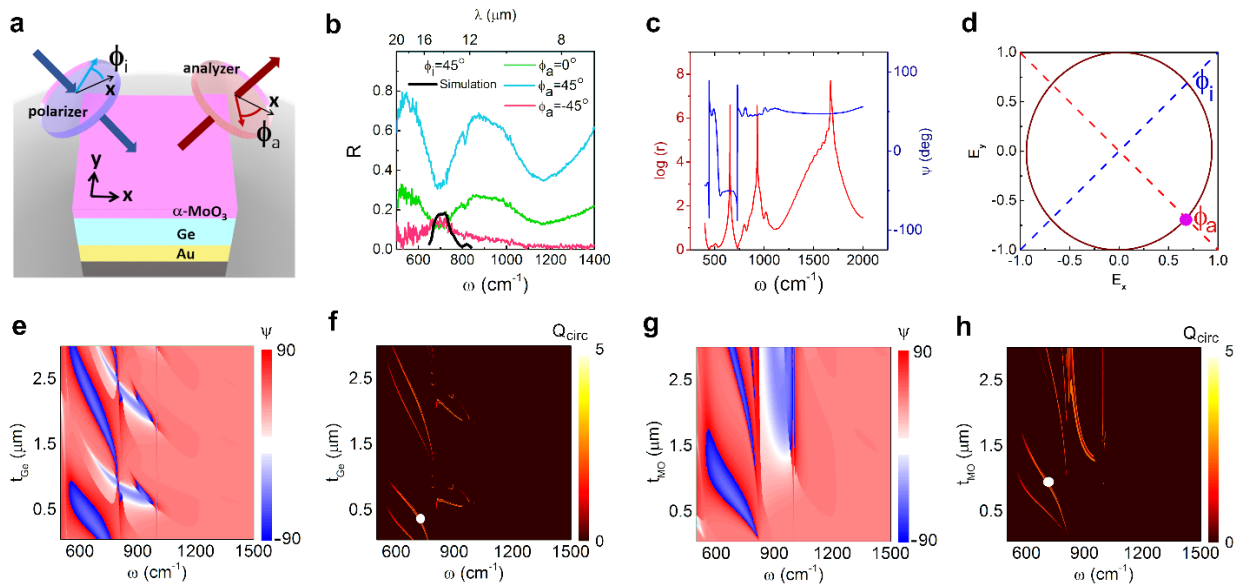

**Figure 5|RS band enhanced birefringence in $\alpha$-MoO$_3$.** (a) Schematic of depolarization experiment, (b) reflectance of sample S2 for different analyzer values $\phi_a$ when polarizer is set to $\phi_i$ = 45°, the black curve is the simulated intensity for $\phi_a$ = -45°. (c) Simulated polarization ellipse characteristics, ellipticity (*r*) and polarization ellipse major axis angle ($\psi$). (d) Simulated polarization ellipse of reflected beam at $\omega$ = 727 cm$^{-1}$ which represents circularly polarized light. The purple dot represents an example of the |**E**| value from the reflected polarization ellipse at $\omega$ = 727 cm$^{-1}$ used for obtaining the black simulation curve in panel (b). (e) Spectral polarization ellipse major axis angle ($\psi$) versus Ge thickness ($t_{Ge}$) and (f) spectral quarter-wave plate action quality factor ($Q_{circ}$) versus $t_{Ge}$, with $\alpha$-MoO$_3$ thickness $t_{MO}$ = 0.95 μm and

incident angle $\theta$ = 25° and $\phi_i$ =45°. (g) $\psi$ versus $t_{MO}$ and (h) $Q_{circ}$ versus $t_{MO}$, with $t_{Ge}$ = 0.42 µm and $\theta$ = 25° and $\phi_i$ = 45°. The white dots in panels (f) and (h) are the experiment data points representing the circularly polarized reflected beam in panel (d).

Given the in-plane hyperbolicity of $\alpha$-MoO$_3$ between x and y phonon resonances, we anticipate a different polarization state for the reflected beam. In Fig. 5b, when $\phi_i$ = $\phi_a$ = 45°, we observe the reflectance pattern similar to Fig. 4b (R = 1 – A) except that the intensity is lower since a proportion of the power is lost as the reflected beam filters out any polarization other than $\phi_a$. Keeping $\phi_i$ = 45° and modifying to $\phi_a$ = 0° results in decreased reflected intensity, which is expected, again due to filtering of the polarization. The most interesting scenario unfolds when analyzer is cross-polarized to have 90° phase difference with the incident beam ($\phi_i$ = 45°, $\phi_a$ = -45°). If the material were isotropic, the reflected cross-polarized signal would vanish, however, there is reflected intensity that peaks at 727 cm$^{-1}$ as high as 15% from this sample. This infers significant depolarization taking place in the sample and justifies significant polarization rotation associated with the in-plane anisotropy within the lowest frequency RS band. Whilst in principle the addition of a quarter wave plate would allow us to characterize the polarization state of this reflected beam, accurate quarter wave plates are not commercially available covering this spectral range. Therefore, we rely on FDTD simulations, replicating the experimental configuration, to shed light on the spectral depolarization characteristics of the reflected beam.

Sample S2 has been simulated using FDTD in identical experiment conditions to show that our cavity design supports a range of different polarization states and polarization ellipse rotations, specifically, circularly polarized light from a linearly polarized incident beam. The far-field polarization ellipse characteristics of the structure for reflected beam is extracted and fitted to the ellipse equation, $E_x/E_{0x}^2 + E_y/E_{0y}^2 - 2(E_xE_y/E_{0x}E_{0y})\cos(\delta) = \sin^2(\delta)$, where $\delta = \delta_y - \delta_x$ is the phase difference between the phases of electric field vectors in y and x directions and $2E_{0x}$ and $2E_{0y}$ are the major and minor axes of the polarization ellipse. Circularly polarized reflectance requires the phase difference between $E_x$ and $E_y$ ($\delta$) to be 90° and their amplitude ratio to be 1 to satisfy a circle equation. Figure 5c illustrates the logarithmic scale of the ellipticity ($r = E_{0y}/E_{0x}$), i.e. the ratio of major to minor axes of the polarization ellipse (blue curve). In this plot, spikes represent linearly polarized light, where the logarithmic ratio is presented; otherwise, the spikes at linear polarization points would compromise the legibility of other data points. In particular, this logarithmic plot approaches zero at $\omega$ = 727 cm$^{-1}$, representing almost equal $E_{0x}$ and $E_{0y}$ amplitudes with exact ratio of $r$ = 1.066. The red curve in Fig. 5c represents the angle of major axis of the polarization ellipse with respect to x axis ($\psi$). The step near $\omega$ = 730 cm$^{-1}$ for this curve has a significant implication; the polarization ellipse and hence field amplitudes extend for both axes, until near 727 cm$^{-1}$, the major axis is abruptly changed from values close to 45° to the ones near -45°. During this transition, there exists a frequency at which major and minor axes are equal in amplitude and the phase difference between $E_x$ and $E_y$ approaches 90°. This happens at 727 cm$^{-1}$ where $\delta_y - \delta_x$ = -89.16°. Thus, at $\omega$ = 727 cm$^{-1}$, α-MoO$_3$ changes the linearly polarized incident light to a circularly polarized reflected beam, the polarization ellipse of which is illustrated in Fig. 5d, representing left-handed circularly polarized light. In order to draw a connection between simulation and experiment we record the spectral ratio of the total electric field in $\phi_a$ = -45° direction (dashed red line in Fig. 5d) to the electric field of the incident intensity ($\phi_i$ = 45°, blue dashed line in Fig. 5d) from the simulations. In other words, the magnitude of the normalized electric field (|**E**|) is recorded in the intersection of reflected polarization ellipse with analyzer $\phi_a$ = - 45° at several frequency points (purple dot in Fig. 5d is an example at $\omega$ = 727 cm$^{-1}$). Afterwards, the normalized field value is squared to represent intensity and is multiplied to the spectral reflectance curve measured for sample S2 in Fig. 4b when $\phi_i$ = 45° (R = 1 – A, where A is the dark green curve in Fig 4b for $\phi_i$ = 45°). The result is the black curve in Fig. 5b which follows the peak trend of the experimental pink curve. Figure 5b shows that 15% of the incident beam is converted to cross-polarized state. Hence, linearly polarized incident light is modified to circularly polarized reflected light at $\omega$ = 727 cm$^{-1}$ from measurements hence exhibiting quarter-wave plate action at IR frequencies.

Quarter-wave plate frequency can be tuned by adjusting the thickness of layers in this FP structure. With the aim of obtaining circularly polarized radiation in the reflected beam, the spectal polarization ellipse major axis angle ($\psi$) versus Ge thicness ($t_{Ge}$) and α-MoO$_3$ thickness ($t_{MO}$) are provided respectively in Figs. 5e and g. The transformation of $\psi$ between positive and negative values marks a transformation that can support quarter-wave plate action. The transitions lay between the blue and red regions in Figs. 5e and g. A quality factor can be defined to track the behavior of circular polarization as follows,

$$Q_{circ} = \ln(|\frac{1}{\frac{\pi}{2}-|\delta|+(1-r)}|). \qquad (2)$$

Equation (2) takes the two conditions of circularly polarized light into account; $Q_{circ}$ becomes very large when the phase difference between $E_y$ and $E_x$ is 90° (|$\delta$| = $\pi/2$) and magnitude ratio of the two in-plane components of the fields is 1 ($r$ = 1). The logarithmic scale is used to improve the clarity of the resulting acceptable data points. In order to single out the most reliable conversions, $\delta$ and $r$ are only allowed to deviate from the ideal circularly polarized case by an error margin of 5%. The results for sweeping $t_{Ge}$ and

$t_{MO}$ are respectively demonstrated in Figs. 5f and h which are in agreement with the previously simulated results with FDTD full-wave software in Supplementary Figures 6c and d. As observed earlier, the lines in Figs. 5f and h are all located within the borders of the negative and positive $\psi$ transformations in Figs. 5e and g. We also note that significant depolarization to states other than circular also takes place in this polarization conversion. The mentioned discrepancy is attributed to the rough surface of $\alpha$-MoO$_3$ flakes, minimization of which is canonical for efficient polarization conversion applications. In the case of smooth $\alpha$-MoO$_3$ flakes, using simulated reflectance curve from sample S2 for $\phi_i$ = 45° instead of the experimental one, we predict conversion efficiency of over 90% as the ideal limit where all the non-ideal effects are absent. The observed reflectance in experiments, however, suggests a combination of different polarizations that obstructs high efficiency for single polarization in the output. Besides, other polarization rotation and states are also attainable by this configuration. As an example, the ellipticity demonstrates spikes (large *r*) which represent linearly polarized reflected beam. At $\omega$ = 655 cm$^{-1}$, *r* = 736 and $\psi$ = -50.4° which infers that for the reflected beam, linear polarization state of the incident beam is maintained and is rotated by -95.4°.

The flake thicknesses required for quarter and half-wave plate actions are more than an order of magnitude smaller than the wavelength of interest. OPhs render the refractive index of $\alpha$-MoO$_3$ dispersive. The limit for thinner flakes for polarization conversion action can be pushed further if the frequency of operation is set closer to the OPh resonances. In such a scenario, however, there exists a trade-off between thin flake and loss of power since OPhs absorb light. It is worth pointing out that for an $\alpha$-MoO$_3$ layer on Ge substrate (*i.e.* absence of Au layer), the required $\alpha$-MoO$_3$ thickness values are two times higher compared to the FP structure proposed in this study. Additionally, due to the existence of transmittance in the absence of Au layer, it is more challenging to satisfy the amplitude condition of half-wave plate action which sets more limitations on the accessible frequency and thickness values for circularly polarized radiation. The reflected power in the absence of Au layer is at least 15% less which demonstrates the significance of FP structure for higher efficiency, thinner requirement for samples and the availability of more thickness parameters for tuning ($t_{MO}$ and $t_{Ge}$ in contrast to only $t_{MO}$ for the structure in the absence of Au). The polarization characteristics can also be tuned with the incident polarization angle, which are provided in Supplementary Note 3 along with the results for $\alpha$-MoO$_3$-Ge structure.

It is worth pointing out that in-plane anisotropic 2D materials have been previously investigated for birefringence and dichroism. Particularly, anisotropic plasmons observed in black phosphorus (BP) were examined to modify the polarization of the reflected or transmitted beam at IR frequencies. However, adding a substrate proved to be detrimental for efficiency and the relevant studies were restricted to simulations of patterned free-standing BP. Moreover, unlike $\alpha$-MoO$_3$, BP is unstable in the ambient and needs to be capped with dielectrics which affect the efficiency. The highest reported reflected 90° polarization rotation reported in simulation studies are in the order of 10% [33,34]. Here, without any film patterning, we have a true experimental demonstration of 90° polarization rotation (Supplementary Note 3) in reflectance mode at IR frequencies which reaches to measured value of 12% reflectance in this study and is comparable to the theoretical reports with patterned anisotropic plasmonic materials, while we anticipate a theoretical value of 90% for ideal and smooth flakes. This experiment emphasizes the potential of $\alpha$-MoO$_3$ for modifying the polarization of radiation in the IR obviating the need for patterning.

**Discussion.**

In this work we have demonstrated the polarization-dependent optical responses of $\alpha$-MoO$_3$ FP cavities numerically and experimentally. $\alpha$-MoO$_3$ is a vdW semiconductor with strong longitudinal and transverse optical phonon resonances that give rise to high-quality absorption peaks in all the three orthogonal directions distinctly. In order to verify this, $\alpha$-MoO$_3$ flakes were transferred to a multilayer cavity structure composed of Au and Ge. FDTD simulations and TMM method was used to analyze the structure and dispersion relations were obtained by scrutinizing the contribution of each layer to the absorption/reflection characteristics. The observed strong anisotropy in the simulations were confirmed experimentally that supported the existing strong birefringence. By engineering the layer thicknesses, on-resonance absorption peaks of over 80% were observed for OPh$_x$ and quality factors as high as 119 were verified experimentally for OPh$_z$. Afterwards, the off-resonance RS band enhanced in-plane birefringence of $\alpha$-MoO$_3$ was demonstrated experimentally with corroborating simulations that showed 15% reflectance of the cross-polarized state. This has implications for IR photonics as $\alpha$-MoO$_3$ naturally enables modification of the amplitude, phase and polarization state of radiation essentially eliminating the need for patterning and further processing. It was also demonstrated that tailoring mode engineering through the tuning of the thickness of layers in FP configuration, polarization filtering and rotating frequency can be tuned for on-demand photonic applications.

Given the growing interest in PhPs as low-loss polaritons for photonic applications at IR wavelengths, $\alpha$-MoO$_3$ is a valuable addition to the family of vdW materials. Despite having access to high-quality phonon resonances in hBN, there remain restrictions on IR photonics since hBN is uniaxial and an insulator.

Furthermore, for all PhP-supporting materials, the frequency domain over which the RS band is realized is dictated by the optic phonon frequencies, with minimal tuning possible. However, integration of free-carrier injection for a semiconducting PhP material offers the potential to leverage the LO-phonon-plasmon-coupling (LOPC) effect to induce spectral tuning of the PhP modal frequencies[45]. The semimetal graphene and semiconductor black phosphorus have been confirmed respectively as plasmonic and anisotropic plasmonic materials for IR photonics. The addition of vdW semiconducting $\alpha$-MoO$_3$ with biaxial properties can relax several restrictions in photonic device design and combined with the existing wealth of IR vdW actively tunable plasmonic and phononic materials, paves the path to the next generation of switchable photonic integrated devices for the less investigated IR frequencies. We envisage that high-quality $\alpha$-MoO$_3$ layers can achieve large-area and efficient polarization-sensitive optical components in IR as the fabrication techniques mature to provide large-area, smooth $\alpha$-MoO$_3$ films.

## Methods.

### Multilayer structure fabrication

The multilayer FP structure was fabricated using physical vapor deposition (PVD) techniques. Au was deposited using AJA eBeam Evaporator system and the base and deposition pressures were respectively $2\times10^{-6}$ and $9\times10^{-6}$ Torr. The deposition rate was 0.5 nm/s. Ge was similarly deposited using Auto eBeam evaporation system with base and deposition pressure of $2\times10^{-6}$ and $5\times10^{-6}$ Torr. The deposition rate of this sublime material was held within 0.13 and 0.17 nm/s during the deposition.

### $\alpha$-MoO$_3$ fabrication and transfer

$\alpha$-MoO$_3$ flakes were grown using low pressure PVD. For this process, 50 mg of MoO$_3$ (Sigma-Aldrich) powder was spread evenly within an alumina boat. This boat was placed within a 1-inch diameter quartz tube and at the center of a small Lindberg tube furnace. A 41 inch$^2$ rectangular piece of SiO$_2$/Si wafer (300 nm oxide thickness) was placed face-up downstream in a colder zone of the furnace. These pieces were suspended on top of an alumina boat and were located roughly 4 cm from the center region. The pressure was maintained at 2.8 Torr with a carrier gas of O$_2$ at a flow rate of 25 sccm. The center of the furnace was then heated to 675 °C over a period of 25 minutes and then to 700 °C over a period of 5 minutes. Upon reaching 700 °C, the furnace was immediately opened, thereby quenching the deposition.

A polycarbonate-based process was used to transfer the deposited flakes from the SiO$_2$/Si substrate to several different substrates. This process involved coating the SiO$_2$/Si substrate with a polycarbonate solution (5% polycarbonate to 95% chloroform by weight) at 2000 rpm for 60 seconds and baking at 120 °C for 1 minute. The substrate was then placed in water to allow the polycarbonate film and deposited $\alpha$-$MoO_3$ flakes to naturally delaminate from the film. The polycarbonate film was then removed from the water and allowed to dry. Once dry, the polycarbonate film was placed onto the target substrate and baked at 170 °C for 15 minutes. Finally, the target substrate was soaked overnight in chloroform and rinsed in isopropyl alcohol immediately thereafter to dissolve the polycarbonate film and leave only the $\alpha$-$MoO_3$ flakes.

**FTIR characterization**

Mid-infrared reflectance measurements were taken with Hyperion 2000 IR microscope coupled to a Bruker Vertex 70 FTIR spectrum. For the FP structures MCT detector with KRS5 polarizer was used. The Cassegrain objective was 36x and the aperture dimensions were 50 by 50 $\mu m^2$ and the spectra were taken with 22.5° polarization steps. The polarizers used for the depolarization experiments were KRS5 wire grid polarizer from pike technologies and HDPE wire grid polarizer.

**Additional Information**

Supplementary information redacted.

# References.

1. Amako, J., & Sonehara, T. Kinoform using an electrically controlled birefringent liquid-crystal spatial light modulator. *Appl. Opt.* **30**, 4622-4628 (1991).
2. Xun, X., & Cohn, R. W. Phase calibration of spatially nonuniform spatial light modulators. *Appl. Opt.* **43**, 6400-6406 (2004).
3. Badoz, J., Billardon, M., Canit, J. C. & Russel, M. F. Sensitive devices to determine the state and degree of polarization of a light beam using a birefringence modulator. *J. Opt.* **8**, 373 (1977).
4. McIntyre, C. M., & Harris, S. E. Achromatic wave plates for the visible spectrum. *JOSA* **58**, 1575-1580 (1968).
5. Yang, H., *et. al.* Optical waveplates based on birefringence of anisotropic two-dimensional layered materials. *ACS Photon.* **4**, 223-3030 (2017).
6. Saleh, B. E. A., & Teich, M. C. *Fundamentals of photonics.* (John Wiley & Sons, 2019).
7. Rodrigues, M. J. L. F. Resonantly increased optical frequency conversion in atomically thin black phosphorus. *Adv. Mater.* **28**, 10693-10700 (2016).
8. Ren, M., Plum, E., Xu, J. & Zheludev, N. I. Giant nonlinear optical activity in a plasmonic metamaterial." *Nat. Commun.* **3**, 1-6 (2012).
9. De Boer, J. F., Milner, T. E., Van Gemert, M. J. C. & Stuart Nelson, J. Two-dimensional birefringence imaging in biological tissue by polarization-sensitive optical coherence tomography. *Opt. Lett.* **22**, 934-936 (1997).

10. Low, T. *et. al.* Polaritons in layered two-dimensional materials. *Nat. Mater* **16**, 182 (2017).
11. Correas-Serrano, D., Alù, A., & Gomez-Diaz, J. S. Plasmon canalization and tunneling over anisotropic metasurfaces. *Phys. Rev. B* **96**, 075436 (2017).
12. Folland, T. G. & Caldwell, J. D. Precise control of infrared polarization using crystal vibrations. *Nature*, 499–501 (2018).
13. Dai, S. *et. al.* Subdiffractional focusing and guiding of polaritonic rays in a natural hyperbolic material. *Nat. Commun.* **6**, 6963 (2015).
14. Poddubny, A., Iorsh, I., Belov, P., & Kivshar, Y. Hyperbolic metamaterials. *Nat. Photon.* **7**, 948 (2013).
15. Li, Peining, Martin Lewin, Andrey V. Kretinin, Joshua D. Caldwell, Kostya S. Novoselov, Takashi Taniguchi, Kenji Watanabe, Fabian Gaussmann, and Thomas Taubner. Hyperbolic phonon-polaritons in boron nitride for near-field optical imaging and focusing. *Nat. Commun.* **6**, 1-9 (2015).
16. Shi, K., Bao, F., & He, S. Enhanced near-field thermal radiation based on multilayer graphene-hBN heterostructures. *ACS Photon*. **4**, 971-978 (2017).
17. Hajian, H., Ghobadi, A., Butun, B., & Ozbay, E. Nearly perfect resonant absorption and coherent thermal emission by hBN-based photonic crystals. *Opt. Express* **25**, 31970-31987 (2017).
18. Lin, X. *et. al.* All-angle negative refraction of highly squeezed plasmon and phonon polaritons in graphene–boron nitride heterostructures. *PNAS.* **114**, 6717-6721 (2017).
19. Wu, C., *et. al.* Spectrally selective chiral silicon metasurfaces based on infrared Fano resonances. *Nat. Commun.* **5**, 1-9 (2014).
20. Wadsworth, S. L. & Boreman, G. D. Broadband infrared meanderline reflective quarter-wave plate. *Opt. Express* **19**, 10604-10612 (2011).
21. Song, X., Liu, Z., Xiang, Y., & Aydin, K. Biaxial hyperbolic metamaterials using anisotropic few-layer black phosphorus. *Opt. Express* **26**, 5469-5477 (2018).
22. Song, X., Liu, Z., Scheuer, J., Xiang, Y., & Aydin, K. Tunable polaritonic metasurface absorbers in mid-IR based on hexagonal boron nitride and vanadium dioxide layers. *J. Phys. D* **52**, 164002 (2019).
23. Tran, T. T., Bray, K., Ford, M. J., Toth, M., & Aharonovich, I. Quantum emission from hexagonal boron nitride monolayers. *Nat. Nanotechnol.* **11**, 37 (2016).
24. Korzeb, K., Gajc, M. & Pawlak, D. A. Compendium of natural hyperbolic materials. *Opt. Express* **23**, 25406-25424 (2015).
25. Grigorenko, A. N., Polini, M., & Novoselov, K. S. Graphene plasmonics. *Nat. Photon.* **6**, 749 (2012).
26. A.K. Geim and I.V. Grigorieva. Van der Waals heterostructures. *Nature* **499**, 419 (2013).
27. Foteinopoulou, S., Devarapu, G. C. R., Subramania, G. S., Krishna, S., & Wasserman, D. Phonon-polaritonics: enabling powerful capabilities for infrared photonics. *Nanophotonics*, **8**, 2129-2175 (2019).
28. Caldwell, J. D., Lindsay L., Giannini V., Vurgaftman I., Reinecke T. L., Maier S. A., & Glembocki O. J. Low-loss, infrared and terahertz nanophotonics using surface phonon polaritons. *Nanophotonics* **4**, 44-68 (2015).
29. Ozbay, E. Plasmonics: merging photonics and electronics at nanoscale dimensions. *Science* **311**, 189-193 (2006).
30. Li, P. *et. al.* Infrared hyperbolic metasurface based on nanostructured van der Waals materials. *Science*, **359**, 892-896 (2018).
31. Cheng, H. Dynamically tunable broadband mid-infrared cross polarization converter based on graphene metamaterial. *Appl. Phys. Lett.* **103**, 223102 (2013).

32. Yang, H. *et. al.* Optical waveplates based on birefringence of anisotropic two-dimensional layered materials. *ACS Photon.* **4**, 3023-3030 (2017).
33. Kotov, O. V., & Lozovik, Yu E. Enhanced optical activity in hyperbolic metasurfaces. *Phys. Rev. B* **96**, 235403 (2017).
34. Khaliji, K., Fallahi, A., Martin-Moreno, L., & Low, T. Tunable plasmon-enhanced birefringence in ribbon array of anisotropic two-dimensional materials. *Phys. Rev. B* **95**, 201401 (2017).
35. Liu, Z., & Aydin, K. Localized surface plasmons in nanostructured monolayer black phosphorus. *Nano Lett.* **16**, 3457-3462 (2016).
36. Ma, W. *et. al*. In-plane anisotropic and ultra-low-loss polaritons in a natural van der Waals crystal. *Nature* **562**, 557 (2018).
37. Zheng, Z. *et. al*. Highly confined and tunable hyperbolic phonon polaritons in van der Waals semiconducting transition metal oxides. *Adv. Mater.* **30**, 1705318 (2018).
38. Zheng, Z. *et. al*. A mid-infrared biaxial hyperbolic van der Waals crystal. *Sci. Adv.* **5**, eaav8690 (2019).
39. Álvarez-Pérez, Gonzalo, et al. "Infrared permittivity of the biaxial van der Waals semiconductor $\alpha$-MoO$_3$ from near-and far-field correlative studies." *Adv. Mater.* 1908176 (2020).
40. Wei, C. *et. al*. Polarization Reflector/Color Filter at Visible Frequencies via Anisotropic α-MoO$_3$. *Adv. Opt. Mater.* (2020).
41. Dai, S. *et. al*. Graphene on hexagonal boron nitride as a tunable hyperbolic metamaterial. *Nat. Nanotechnol.* **10**, 682 (2015).
42. Deinzer, G. & Strauch D. Two-phonon infrared absorption spectra of germanium and silicon calculated from first principles. *Phys. Rev. B* **69**, 045205 (2004).
43. Murthy, A. A. *et. al.* Intrinsic transport in 2D heterostructures mediated through h-BN tunneling contacts. *Nano Lett.* **18** 2990-2998 (2018).
44. Hanson, E. D. *et. al*. Systematic study of oxygen vacancy tunable transport properties of few-layer MoO$_{3-x}$ enabled by vapor-based synthesis. *Advanced Funct. Mater.* **27**, 1605380 (2017).
45. Dunkelberger, A. D., *et. al.* Active tuning of surface phonon polariton resonances via carrier photoinjection. *Nat. Photon.* **12**, 50-56 (2018).

## Acknowledgements.

K.A. acknowledges support from the Office of Naval Research Young Investigator Program (ONR-YIP) Award (N00014-17-1-2425). The program manager is Brian Bennett. K.A and V.P.D. acknowledges partial support from the Air Force Office of Scientific Research under Award Number FA9550-17-1-0348. This material is partially supported by the National Science Foundation under Grant No. DMR-1929356. J.D.C. acknowledges support from the Office of Naval Research under award number (N00014-18-1-2107), while T.G.F. acknowledges support from Vanderbilt University School of Engineering. A.A.M. gratefully acknowledges support from the Ryan Fellowship and the IIN at Northwestern University. This work made use of the EPIC, Keck-II, SPID, and Northwestern University Micro/Nano Fabrication Facility (NUFAB) facilities of Northwestern University's *NUANCE* Center, which has received support from the Soft and Hybrid Nanotechnology Experimental (SHyNE) Resource (NSF ECCS-1542205); the MRSEC program (NSF

DMR-1720319) at the Materials Research Center; the International Institute for Nanotechnology (IIN); the Keck Foundation; and the State of Illinois, through the IIN.## Author Contributions.

S.AD., X. S. and K.A. conceived the idea, S.AD. designed the devices and conducted the fabrication, experiments, measurements and simulations and prepared the paper. T.G.F. designed the depolarization experiment, conducted low frequency FTIR experiments and helped in direction and writing of the paper. A.M. contributed in the growth, transfer and characterization of $\alpha$-MoO$_3$ flakes. X.S. took part in design, measurements and FP simulations. I.T. contributed in the simulations of depolarization. All authors commented on the manuscript and V.P.D., J.D.C. and K.A. supervised the project.